\documentclass[11pt,a4paper]{article}
\usepackage[hyperref]{paper}
\usepackage{times}
\usepackage{latexsym}

\usepackage{url}

\usepackage{graphicx}
\usepackage{booktabs}
\usepackage{mathtools} 
\usepackage{multirow} 
\usepackage{tabularx} 
\usepackage{subfig}
\usepackage{tablefootnote}
\usepackage{placeins}
\usepackage{amssymb}
\usepackage{enumitem}
\usepackage{amsmath}
\usepackage{stfloats}

\newcommand\blfootnote[1]{%
  \begingroup
  \renewcommand\thefootnote{}\footnote{#1}%
  \addtocounter{footnote}{-1}%
  \endgroup
}

\aclfinalcopy 

\newcolumntype{P}[1]{>{\centering\arraybackslash}p{#1}}

\title{Multivariate Realized Volatility Forecasting with Graph Neural Network}

\author{Qinkai Chen$^{\dag}$$^{\ddag}$ \, Christian-Yann Robert$^{\mathsection}$\\ 
$^{\dag}$Ecole Polytechnique, Palaiseau, France \\
$^{\ddag}$Exoduspoint Capital Management France, Paris, France \\
$^{\mathsection}$ENSAE Paris, Palaiseau, France \\
\tt{qinkai.chen@polytechnique.edu} \\
\tt{christian-yann.robert@ensae.fr}
}

\date{}

\begin{document}

\maketitle

\begin{abstract}
  The existing publications demonstrate that the limit order book data is useful in predicting short-term volatility in stock markets.
  Since stocks are not independent, changes on one stock can also impact other related stocks.
  In this paper, we are interested in forecasting short-term realized volatility in a multivariate approach based on limit order book data
  and relational data. To achieve this goal, we introduce Graph Transformer Network for Volatility Forecasting. 
  The model allows to combine limit order book features and
  an unlimited number of temporal and cross-sectional relations from different sources. 
  Through experiments based on about 500 stocks from S\&P 500 index, we find a
  better performance for our model than for other benchmarks.
\end{abstract}

\section{Introduction}
\label{sec:introduction}

\blfootnote{The authors would like to thank Mathieu Rosenbaum from Ecole Polytechnique for his valuable 
guidance and advice during this work.
The authors also would like to express our gratitude to Kaggle users AlexiosLyon and Max2020 for their posts and discussions.
The authors also appreciate the insightful discussions with Jianfei Zhang.}

Volatility is an important quantity in finance, it evaluates the price fluctuation and represents
the risk level of an asset. It is one of the most important indicators used in risk management
and equity derivatives pricing. Although the volatility is not observable, \citet{andersen1998answering}
show that realized volatility is a good estimator of volatility.
Forecasting realized volatility has therefore attracted the attention of various researchers.

\citet{brailsford1996evaluation} propose using GARCH (Generalized AutoRegressive Conditional Heteroskedasticity) models
to forecast realized volatilities based on daily prices. \citet{gatheral2010zero} introduce several
simple volatility estimators based on Limit Order Book (LOB) data, showing that the use of LOB data
can lead to better predicting results. More recently, \citet{rahimikia2020machine} and \citet{zhang2020universal}
use machine learning techniques such as Recurrent Neural Network (RNN) to improve such predictions.

The aforementioned literatures adopt a univariate approach in this task, which means that
the model only considers one outcome for one stock at the same time, instead of jointly
considering the situations of all stocks, although the asset returns on the financial markets can
be highly correlated \citep{campbell1993trading}.
\citet{andersen2005volatility} propose linear multivariate volatility forecasting methods based on daily price data
to take into account this correlation, while \citet{bucci2020cholesky} further uses neural networks to forecast realized volatility covariance matrix non-linearly.
\citet{bollerslev2019high} first propose a parametric multivariate model based on LOB data using covolatility and covariance
matrices. More recently, a Kaggle multivariate realized volatility prediction competition\footnote{https://www.kaggle.com/c/optiver-realized-volatility-prediction} 
was sponsored by Optiver to challenge data scientists to propose new multivariate forecasting methods.

Compared with the univariate approach, multivariate models can capture the relations among observations.
Most recently, Graph Neural Networks (GNN) \citep{bruna2013spectral} are proposed to integrate such relationship
into the commonly used non-linear neural networks.
This approach achieves significant success in multiple applications, such as traffic flow prediction \citep{li2017diffusion},
recommender systems \citep{berg2017graph} and stock movement prediction \citep{sawhney2020deep,chen2021graph}.
To the best of our knowledge, no graph-based structure for volatility forecasting has been proposed in the literatures.

Hence, to further improve the volatility forecasting performance, inspired by previous researches (\textbf{Sec. 2}), the
Kaggle competition and our real-life
use cases, we build a multivariate volatility forecasting model (\textbf{Sec. 3}) based on Graph Neural Network: Graph Transformer
Network for Volatility Forecasting (GTN-VF). This model predicts the short-term volatilities from LOB data and both
cross-sectional and temporal relationships from different sources (\textbf{Sec. 4}). With various experiments on about 500 stocks
from the S\&P 500 index, we demonstrate that GTN-VF outperforms other baseline models with a significant margin on
different forecasting horizons (\textbf{Sec. 5}).

\section{Related Work}

\subsection{Volatility Forecasting}

As introduced in Section \ref{sec:introduction}, there are two types of volatility forecasting models: 
univariate models and multivariate models.

In univariate approaches, researchers can design intuitive estimators \citep{zhou1996high,zhang2006efficient}, without considering
relations among observations.
More commonly, researchers adopt time-series methods to model the time dependence
while ignoring the cross-sectional relations and calibrating one set of parameters for each asset.
For example, \citet{brailsford1996evaluation} propose GARCH models, \citet{sirignano2019universal}
use RNN models with Long Short-Term Memory (LSTM) and \citet{ramos2021multi} adopt a Transformer structure \citep{vaswani2017attention}.

Multivariate models add cross-sectional relationships and achieve better results. 
For example, \citet{kwan2005multivariate} introduce multivariate threshold GARCH model and
\citet{bollerslev2019high} propose a multivariate statistical estimator based on co-volatility matrices.
It is worth noting that all models above only use asset covariance as the source to build the relationship while
ignoring the intrinsic relations between the companies that issue the stocks.

\subsection{Graph Neural Network}
\label{subsec:graph_neural_network}

A graph is composed of nodes and edges, where a node represents an instance in the network and an edge denotes
the relationship between two instances. It is an intuitive structure to describe the relational information.
Recently, many researches focus on generalizing neural networks on graph structure to capture non-linear
interactions among the nodes. 

\citet{bruna2013spectral}
first generalize the Convolutional Neural Network (CNN) on graph-based data, while \cite{kipf2016semi}
propose Graph Convolutional Network (GCN) and \citet{defferrard2016convolutional} introduce ChebNet,
both of which have reduced network complexity and better predictive accuracy.

However, the aforementioned models are required to load all the graph data into the memory at the same time,
making training larger relation networks impossible. \citet{gilmer2017neural} state that
Graph Neural Networks are essentially message passing algorithms.
It means that the model makes decision not only based on one node's observation, but also the information passed from
all other related nodes defined in the format of a graph. Based on this generalization, \citet{hamilton2017inductive}
propose GraphSAGE which allows batch training on graph data.

\citet{shi2020masked} further show that using a Transformer-like operator to
aggregate node features and the neighbor nodes' features gives a better performance than a simple
average such as GraphSAGE or an attention mechanism such as
Graph Attention Network \citep{velivckovic2017graph}.

\noindent
To close the gap in the researches, we propose GTN-VF, which adopts the state-of-the-art
Graph Neural Networks to model the relationships. In addition, GTN-VF allows to integrate
an unlimited number of relational information, including both widely used covariance and
other external relations such as sector and supply chain, which were rarely used in previous researches.

\section{Problem Formulation}
\label{sec:problem_formulation}

We formulate this multivariate volatility forecasting problem as a regression task. 
The goal is to predict the realized volatility vector over the next $\Delta T$ seconds at a given
time $t$ with all previously available data. 

We first define the return of stock $s$ at time $t$ as
\begin{equation}
  r_{s, t} = log(\frac{P_{s, t}}{P_{s, t-1}})
\end{equation}
where $P_{s, t}$ is the last trade price of $s$ at $t$.

We then use $RV_{s, t, \Delta T}$ to denote the realized volatility for stock $s$ between $t$ and $t+\Delta T$, it is defined as:
\begin{equation}
  \label{eq:realized_volatility}
  RV_{s, t, \Delta T} = \sqrt{\sum_{i=t}^{t+\Delta T} r_{s, i}^{2}}
\end{equation}

Previous researches \citep{malec2016semiparametric,rahimikia2020big,rahimikia2020machine}
usually calibrate one model for each stock. This prediction model $f_{s}$ for stock $s$ can be written as:
\begin{equation}
  \widehat{RV_{s, t, \Delta T}} = f_{s}([D_{s, t_{1}},...,D_{s, t_{m}}], \theta)
\end{equation}
where $D_{s, t}$ denotes the limit order book data related to stock $s$ between $t$ and $t-\Delta T'$,
and $t_{1}<...<t_{m}<t$.
$\Delta T'$ is a parameter denoting the backward window used to build features for $t$, while $\theta$ represents the model parameters.

However, as stated in Section \ref{sec:introduction}, the realized volatilities of the stocks are related through their LOBs.
We want to consider this effect in our model and we write our prediction model $g_{0}$ as:
\begin{equation}
  \widehat{RV_{s, t, \Delta T}} = g_{0}(\begin{bmatrix}
    D_{s_{1}, t_{1}} & ... & D_{s_{1}, t_{m}}\\
    ... & ... & ...\\
    D_{s_{n}, t_{1}} & ... & D_{s_{n}, t_{m}}
    \end{bmatrix}, 
    \mathcal{G}_{s}, \theta)
\end{equation}
where $\mathcal{G}_{s}$ is the relationship of stock $s$ with all other stocks $s_{1}, ..., s_{n}$.
It means that our model jointly considers all the features from all other stocks when predicting realized
volatility for stock $s$ and its relationship with other stocks, instead of only taking its own features
into account.

In additional to the relationship among stocks, we can also consider the relationship among the timestamps we predict.
For example, at time $t$, we can check whether the behaviors of the stocks are similar to their behaviors at previous timestamps.
We use $\mathcal{G}_{t}$ to denote this temporal relationship.
Our prediction model $g$ is finally written as:
\begin{equation}
  \label{eq:final_formulation}
  \widehat{RV_{s, t, \Delta T}} = g(\begin{bmatrix}
    D_{s_{1}, t_{1}} & ... & D_{s_{1}, t_{m}}\\
    ... & ... & ...\\
    D_{s_{n}, t_{1}} & ... & D_{s_{n}, t_{m}}
    \end{bmatrix}, 
    \mathcal{G}_{s}, \mathcal{G}_{t}, \theta)
\end{equation}

\section{Graph Transformer Network for Realized Volatility Forecasting}

Our model consists of two main components: a LOB data encoder and a Graph Transformer Network.
The LOB data encoder transforms numerical LOB data into multiple features. It also transforms
categorical information, such as the stock ticker, into a fixed-dimension embedding. It finally
concatenates the numerical features and the embedding for categorical features as the node feature.

The Graph Transformer Network then takes all the node features and the pre-defined relationship information as input.
After training, it will give each node a new meaningful embedding which contains information from 
both LOB data and relational data. With a fully connected layer, we can get the final prediction of 
realized volatility for this node.

\begin{figure*}[h]
  \includegraphics[width=\linewidth]{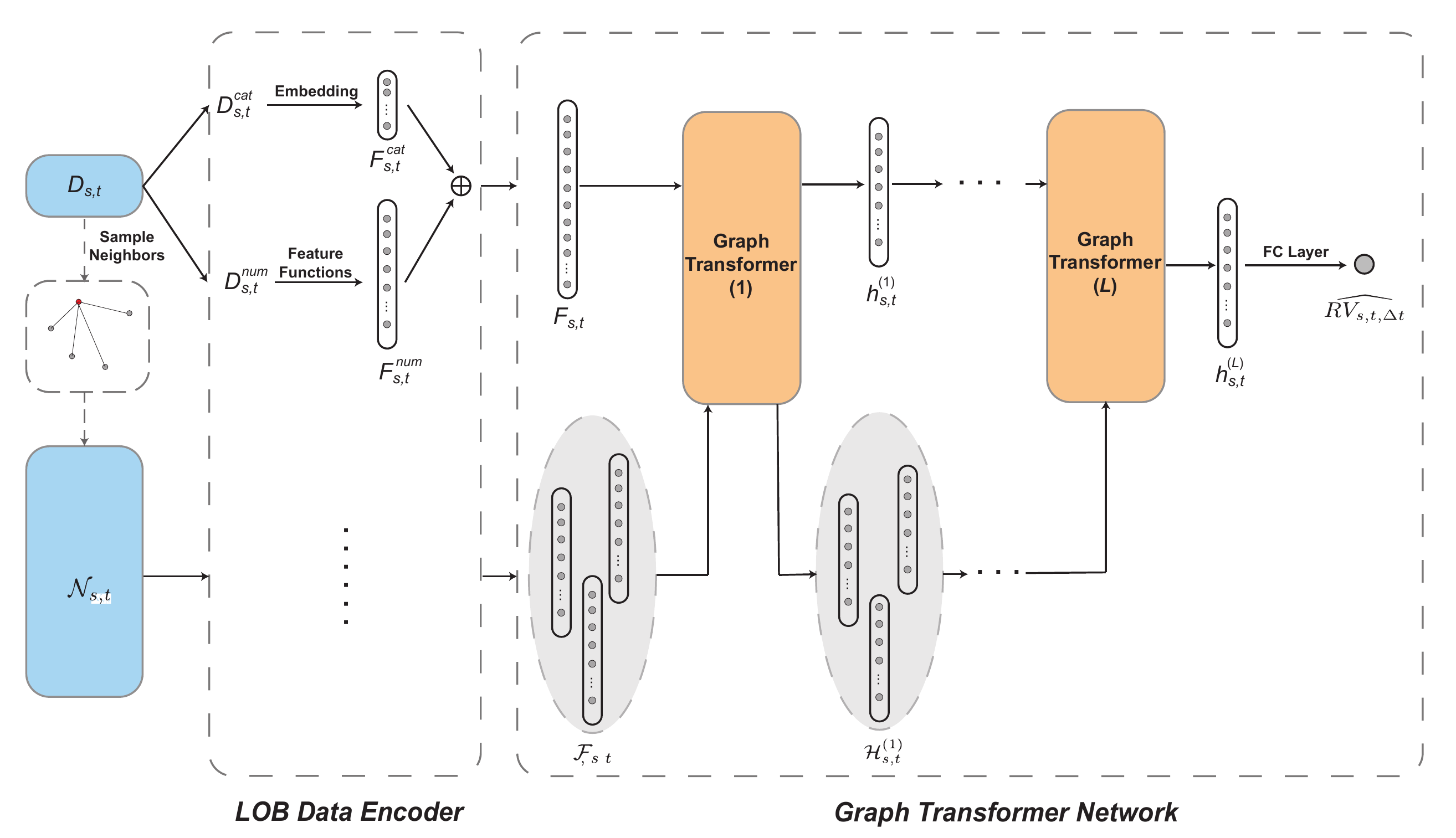}
  \caption{Structure of our Graph Transformer Network for Volatility Forecasting.
  This illustration shows the prediction process for one node $node_{s,t}$. The LOB data encoder first
  transforms its LOB data into a fixed-dimension node feature $F_{s,t}$. GTN then takes this node feature
  and all the features from all other nodes connected with this node ($\mathcal{F}_{s,t}$). After $L$ Graph Transformer operations, 
  we get the node embedding $h^{(L)}_{s,t}$ as output. We then use a fully connected layer to transform this node
  embedding into our final prediction $\widehat{RV_{s, t, \Delta T}}$.}
  \label{fig:gtn_draw}
\end{figure*}

An illustration of the whole structure is shown in Figure \ref{fig:gtn_draw}.

\subsection{LOB data encoder}
\label{subsec:lob_data_encoder}

We first divide our LOB data $D_{s, t}$ into two parts: numerical data $D_{s, t}^{num}$ and categorical data $D_{s, t}^{cat}$.
For numerical data, we can define different functions to aggregate them into numerical values.
Suppose that we have $k_{num}$ such functions $f_{1},...,f_{k_{num}}$, we get $k_{num}$ features
and we put them into a single vector $F_{s, t}^{num}$ where
\begin{equation}
  \label{eq:num_features}
  F_{s, t}^{num} = [f_{1}(D_{s,t}^{num}), ...,f_{k_{num}}(D_{s,t}^{num})]^{\top}
\end{equation}

\noindent
$F_{s, t}^{k_{num}}$ is therefore a vector of size $k_{num} \times 1$.

Following \citet{kercheval2015modelling,bissoondoyal2019asymmetric,makinen2019forecasting}, we define a similar set of
numerical features. We also add some other features which are suitable for our dataset. We show the detailed list
of features we use in our experiments in Appendix \ref{sec:list_node_features}.

For other categorical features, such as stock ticker, we simply adopt an embedding layer to transform them into 
a fixed-dimension vector $F_{s, t}^{cat}$. This vector is of size $k_{cat} \times 1$ where $k_{cat}$ is 
the embedding dimension we can choose.

We then concatenate these two vectors into one vector $F_{s, t} \in \mathbb{R}^{k \times 1}$,
where $k=k_{num} + k_{cat}$.
This operation is written as
\begin{equation}
  \label{eq:node_feature}
  F_{s, t} = F_{s, t}^{num} \oplus F_{s, t}^{cat}
\end{equation}
where $\oplus$ denotes the concatenation operation.

\subsection{Graph Transformer Network}

As stated in Section \ref{subsec:graph_neural_network}, Graph Transformer
operator shows a better performance compared with other structures, we use
it to build our network in this study.

We first build a graph with $m \times n$ nodes where $m$ is the number of timestamps and $n$ is the number of stocks.
Each node $node_{s, t}$ represents the situation of stock $s$ at time $t$. Its initial node feature is $F_{s, t}$ (Equation \ref{eq:node_feature}) encoded by LOB data encoder.
From the relationship $\mathcal{G}_{s}$ and $\mathcal{G}_{t}$, we can find all other nodes connected with $node_{s, t}$. We use $\mathcal{N}_{s, t}$
to denote all the connected nodes. For each stock $s$ at time $t$, the model takes both its own LOB features and the LOB features from other related pairs of $(s,t)$ into account.
It then forecasts the realized volatility based on both self node features and neighbor node features, instead of the traditional approach which
considers only self node features.

The Graph Transformer operator for the $l$-th layer with $C$ heads is written as:
\begin{equation}
  \label{eq:single_head_attention}
  \widehat{h}_{s, t, c}^{(l+1)} = W_{1, c}h_{s,t}^{(l)} + \sum_{node_{i, j} \in \mathcal{N}_{s, t}} \alpha_{i, j, c} W_{2, c} h_{i, j}^{(l)}
\end{equation}
\begin{equation}
  \label{eq:multi_head_aggregation}
  h_{s, t}^{(l+1)} = \sigma (\oplus_{c=1}^{C}\widehat{h}_{s, t, c}^{(l+1)})
\end{equation}

Equation \ref{eq:single_head_attention} first calculates the output vector $\widehat{h}_{s, t, c}^{(l+1)}$ for one single head $c$, in which
$h_{i, j}^{(l)} \in \mathbb{R}^{d_{l} \times 1}$ is the $l$-th layer hidden node embedding for $node_{i, j}$,
$W_{1,c}, W_{2,c} \in (\mathbb{R}^{\widehat{d}_{l+1} \times d_{l}})^{2}$ are trainable parameters. $\alpha_{i, j, c}$ are 
attention coefficients associated with $node_{i,j}$ for head $c$. It is calculated via dot product attention \citep{bahdanau2014neural} by
\begin{equation}
  \alpha_{i,j,c} = softmax(\frac{(W_{3,c}h_{s, t}^{(l)})^{\top} (W_{4,c}h_{i, j}^{(l)})}{\sqrt{d_{l}}})
\end{equation}
where $W_{3,c}$ and $W_{4,c}$ are both trainable parameters of size $\widehat{d}_{l+1} \times d_{l}$.

\begin{figure}[h]
  \includegraphics[width=\columnwidth]{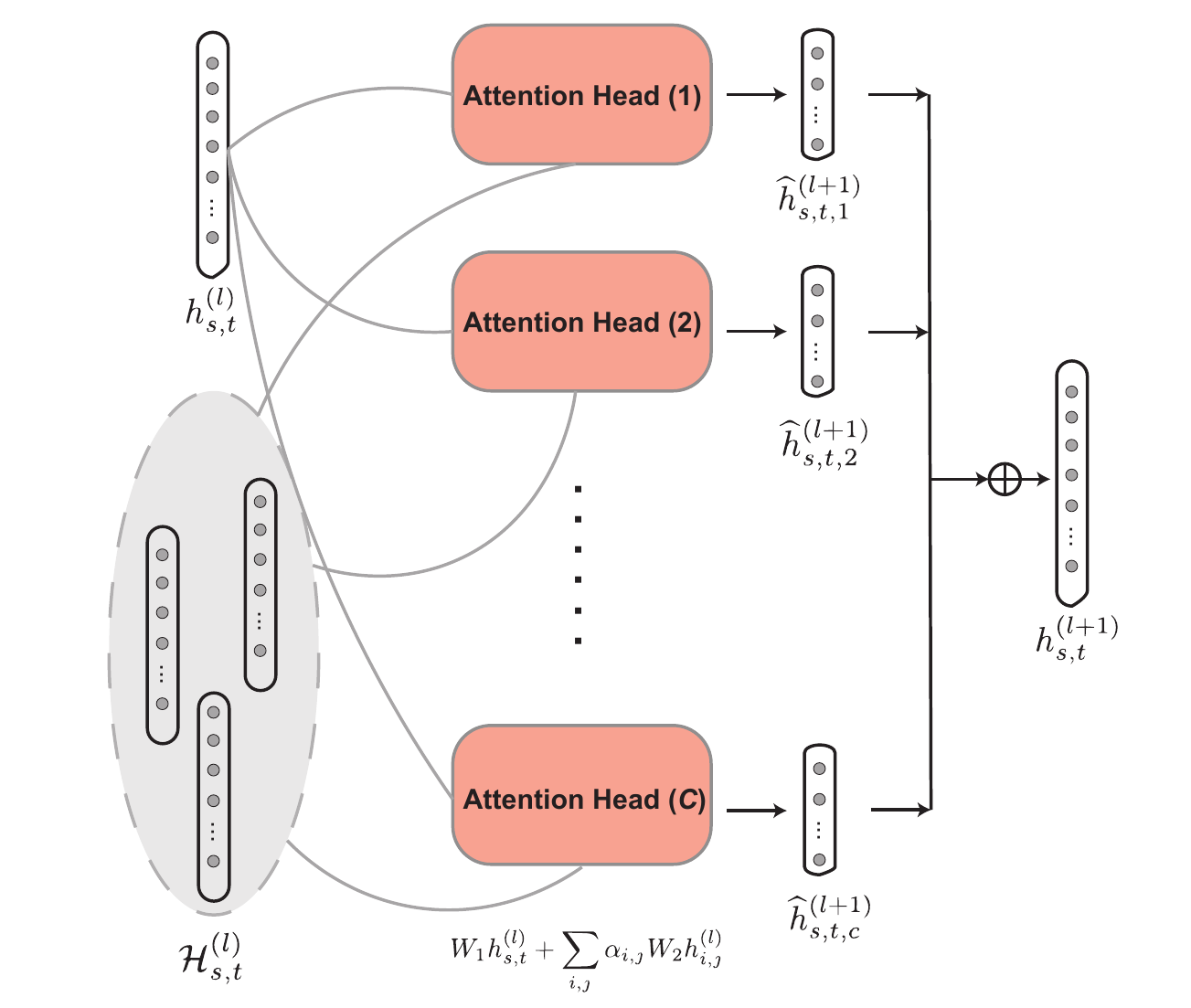}
  \caption{Illustration of Graph Transformer operator. $\mathcal{H}_{s,t}^{l}$ represents the $l$-th layer hidden vectors for all the nodes in $\mathcal{N}_{s,t}$.
   We can then accumulate multiple layers of this structure to build a Graph Transformer Network.}
  \label{fig:graph_transformer_draw}
\end{figure}

We then use Equation \ref{eq:multi_head_aggregation} to aggregate the output from all heads into a final output vector $h_{s, t}^{(l+1)}$ for the $l$-th layer.
It is then used as the input for the $(l+1)$-th layer.
In this equation, $\oplus $ denotes the concatenation operation and $\sigma$ is an activation function such as ReLU \citep{glorot2011deep}.
We show the structure of this operator in Figure \ref{fig:graph_transformer_draw}.

We can then accumulate multiple layers of this structure to better retreive information.
Suppose that our Graph Transformer Network (GTN) has $L$ layers in total, for each node, its initial node features $h_{s, t}^{(0)} = F_{s,t}$
will be transformed into a node embedding $h_{s, t}^{(L)} \in \mathbb{R}^{d_{L} \times 1}$. We then use a fully-connected layer to 
get the final predictions of realized volatility from the node embeddings. This operation is written as
\begin{equation}
  \widehat{RV_{s, t, \Delta T}} = \sigma (W_{0}^{\top}h_{s, t}^{(L)})
\end{equation}
where $W_{0} \in \mathbb{R}^{d_{L} \times 1}$ are trainable parameters in the fully connected layer.

We then use Root Mean Square Percentage Error (RMSPE) as our loss function to evaluate the model and propagate back into the model.
It is define as
\begin{equation}
  \label{eq:rmspe}
  RMSPE = \sqrt{\frac{1}{N} \sum_{s, t} (\frac{\widehat{RV_{s, t, \Delta T}} - RV_{s, t, \Delta T}}{RV_{s, t, \Delta T} + \epsilon})^2}
\end{equation}
where $N$ is the total number of nodes in the graph and $\epsilon$ is a small constant to avoid overflow.

We use RMSPE instead of the standard Mean Square Error (MSE) because the volatilities of different stocks are intrinsically different.
For example, more liquid stocks are usually more volatile than less liquid stocks \citep{domowitz2001liquidity}. 
The RMSPE loss function helps normalize the difference among stocks to make sure that the model has a similar effect on all stocks.

\section{Experiments}

\subsection{LOB data}

We use NYSE daily TAQ data\footnote{https://www.nyse.com/market-data/historical/daily-taq} as our limit order book data.
The data contains all the quotes (only first limit, i.e., best bid and best ask) and trades happening in the US stock exchanges.
We select the entries concerning all the stocks included in the S\&P 500 index\footnote{https://www.spglobal.com/spdji/en/indices/equity/sp-500/\#overview}
as our universe. Compared with other researches (5 stocks in \citet{makinen2019forecasting}, 23 stocks in \citet{rahimikia2020machine}), this large
selection of around 500 stocks also covers some less liquid stocks. We will show that it is more difficult to
have a good prediction on less liquid stocks in Section \ref{subsubsec:accuracy_liquidity}.

\bigskip
\noindent
\textbf{Data Sampling}

Following \citet{barndorff2009realized,rahimikia2020big}, we first sample the data with a fixed frequency $T_{f}$. Since we focus on short-term volatility
forecasting in this paper, we use a \emph{one second} sampling frequency instead of 5 minutes used by \citet{rahimikia2020big} to forecast daily volatility.
Our sampling strategy is as follows:
\begin{itemize}[leftmargin=*,itemsep=0em]
\item For quote data, we snapshot the best ask price ($P_{a}^{1}$), best bid price ($P_{b}^{1}$), best ask size ($V_{a}^{1}$) and best bid size ($V_{b}^{1}$) for each stock
at the end of each second.
\item For trade data, we aggregate all the trades for each stock during each second. We record the number of trades ($N_{t}$), the total number of shares traded ($V_{t}$)
and the volume weighted average price ($P_{t}$)\footnote{$P_{t} = \frac{\sum_{i}^{N_{t}} P_{i}V_{i}}{\sum_{i}^{N_{t}} V_{i}}$ where $P_{i}$ is the price for trade $i$
and $V_{i}$ is the number of shares traded. $N_{t}$ is the total number of trades recorded during second $t$.} of all trades.
\end{itemize}

An example of sampled quote and trade data is shown in Table \ref{tab:example_lob_data}.

\begin{table}
  \scriptsize
  \centering
\subfloat[sampled quote data]{
  \begin{tabular}{ccccccc}
    \toprule
    Date  & Symbol & seconds & $P_{b}^{1}$ & $V_{b}^{1}$ & $P_{a}^{1}$ & $V_{a}^{1}$ \\
    \midrule
    1/3/2017 & A     & 0     & 45.92 & 1     & 46.09 & 1 \\
    1/3/2017 & A     & 1     & 45.92 & 3     & 46    & 2 \\
    1/3/2017 & A     & 2     & 45.92 & 2     & 46    & 2 \\
    1/3/2017 & A     & 5     & 45.92 & 2     & 46    & 1 \\
    1/3/2017 & A     & 6     & 45.94 & 1     & 46.05 & 3 \\
    \bottomrule
    \end{tabular}%
}%
\hfill
\subfloat[sampled trade data]{
  \begin{tabular}{cccccc}
    \toprule
    Date  & Symbol & seconds & $N_{t}$ & $V_{t}$  & $P_{t}$ \\
    \midrule
    1/3/2017 & A     & 0     & 8     & 1500  & 45.802 \\
    1/3/2017 & A     & 1     & 7     & 24959 & 45.94963 \\
    1/3/2017 & A     & 2     & 1     & 300   & 45.96 \\
    1/3/2017 & A     & 3     & 1     & 100   & 45.9544 \\
    1/3/2017 & A     & 5     & 7     & 916   & 45.97817 \\
    \bottomrule
    \end{tabular}%
}
\caption{An example of sampled quote and trade data for stock A on Jan. 3rd, 2017. In addition to the previously
defined LOB fields, we also have \emph{Date}, \emph{Symbol} and \emph{seconds}.
The \emph{seconds} field signifies the 
number of seconds after the market open. We note that the \emph{seconds} are not continuous.
This is because there are occasions that there is no update in the LOB in that second. For quote data it implies that the quote is the same
as the last second, for trade data it implies that there is no trade in that second.}
\label{tab:example_lob_data}%
\end{table}

\bigskip
\noindent
\textbf{Data Bucket}

As introduced in Section \ref{sec:problem_formulation}, our goal is to forecast the realized volatility $\Delta T$ seconds after a given timestamp $t$
based on the features built from a backward window of $\Delta T'$ seconds.

Hence, we need to build buckets which have the length of $\Delta T+\Delta T'$ seconds between $t-\Delta T'$ and $t+\Delta T$. 
In each bucket, we only use the returns between $t$ and $t+\Delta T$
to calculate our target $RV_{s, t, \Delta T}$, we use both returns and other information from LOBs between $t-\Delta T'$ and $t$ to build a set of features 
with LOB data encoder.

In our experiments on US stocks, we create 6 buckets for each stock each day. We select 10:00, 11:00, 12:00, 13:00, 14:00 and 15:00 EST as 6 different $t$.
For the sake of simplicity, we use $\Delta T = \Delta T'$. We use three different $\Delta T$ of 600 seconds, 1200 seconds and 1800 seconds
to show that our model is robust to this choice and is capable of forecasting volatility on different horizons.
This choice also ensures that there is no overlap between buckets to avoid information leakage.
The detailed experiment results are shown in Section \ref{sec:experiment_results}.

\bigskip
\noindent
\textbf{Data Split}

We split our LOB data into 3 parts: train, validation and test.
We ensure that the validation set and the test set are no earlier than the training set to avoid backward looking.
We also remove the buckets where there are no quotes or trades during $\Delta T$.
Detailed statistics of our dataset are shown in Table \ref{tab:data_stats}.

\begin{table}[htbp]
  \small
  \centering
    \begin{tabular}{cccc}
    \toprule
          & train & val   & test \\
    \midrule
    start & Jan-17 & Jan-20 & Jan-21 \\
    end   & Dec-19 & Dec-20 & Oct-21 \\
    \# time & 4,464  & 1,505  & 1,236 \\
    \# stock & 494   & 494   & 494 \\
    \# bucket & 2,141,108 & 743,068 & 607,770 \\
    proportion & 61\%  & 21\%  & 18\% \\
    \bottomrule
    \end{tabular}%
    \caption{The statistics of the LOB data with $\Delta T=600$.}
  \label{tab:data_stats}%
\end{table}%

\subsection{Graph Building}

As stated in Section \ref{sec:problem_formulation}, we consider both temporal ($\mathcal{G}_{t}$) and cross-sectional ($\mathcal{G}_{s}$) relationships among buckets.
In this subsection, we first introduce a method to construct relations without using other data than our LOB data.
We also introduce other relations we constructed with external data, for example, stock sector and supply chain data.

\subsubsection{Temporal Relationship}
\label{subsec:temporal_relationship}

To construct the temporal relationship among the nodes, we only use LOB data.

As introduced in Equation \ref{eq:num_features}, for each $node_{s,t}$, we can calculate $k_{num}$ features.
Suppose that $f_{s, t}^{i}$ is the $i$-th feature for stock $s$ at time $t$. For each time $t$, we let
\begin{equation}
  Q_{t}^{i} = ([f_{1, t}^{i}, ..., f_{n,t}^{i}])^{\top}
\end{equation}
where $Q_{t}^{i} \in \mathbb{R}^{n \times 1}$ represents the feature $i$ of all stocks at time $t$.

Given a time $t_{0}$, we calculate the RMSPE (Equation \ref{eq:rmspe}) of $Q^{i}$ between $t_{0}$ and all other $t$.
We then choose the $K$-smallest RMSPE to form $K$ pairs of times, represented by 
$\mathcal{G}_{t_{0}}=[(t_{0}, t_{1}),...,(t_{0}, t_{K})]$. Then for each such pair $(t_{i},t_{j})$, we connect the two nodes
where $s$ is the same and the time is $t_{i}$ and $t_{j}$ respectively. This can be written as:
\begin{equation}
  \begin{split}
    \forall (t_{i},t_{j}) \in \mathcal{G}_{t_{0}}, \forall k \in [1, n], \\
    \textrm{connect}\ node_{s_{k}, t_{i}}\ \textrm{and}\ node_{s_{k}, t_{j}}
  \end{split}
\end{equation}
After this operation, we get $K \times n$ single-directed edges in the graph for $t_{0}$.
We then repeat the same process for all $t$, and get $K \times n \times m$ edges in total.

In our study, we use two features to get $2 \times K \times n \times m$ edges in the graph,
namely, the average quote WAP\footnote{Weighted Average Price, defined as $\frac{P_{b}^{1} V_{a}^{1} + P_{a}^{1} V_{b}^{1}}{V_{a}^{1} + V_{b}^{1}}$} 
for the first 100 seconds and the average quote WAP for the last 100 seconds in the bucket.

\subsubsection{Cross-sectional Relationship}
\label{subsubsec:cross_sectional_relationship}

Unlike times, there are intrinsic relations among stocks since each stock represents a company in the real life.
Hence, in addition to building relationship based on LOB features, we can also build the cross-sectional graph
with external data.

\bigskip
\noindent
\textbf{Feature Correlation}

Using the same idea introduced in \ref{subsec:temporal_relationship}, we first build
\begin{equation}
  Q_{s}^{'i} = ([f_{s, 1}^{i}, ..., f_{s,m}^{i}])^{\top}
\end{equation}
$Q_{s}^{'i} \in \mathbb{R}^{m \times 1}$ represents the feature $i$ of stock $s$ at $t$.

We then build the edges for $s_{0}$ with
\begin{equation}
  \begin{split}
    \forall (s_{i},s_{j}) \in \mathcal{G}_{s_{0}}, \forall k \in [1, m], \\
    \textrm{connect}\ node_{s_{i}, t_{k}}\ \textrm{and}\ node_{s_{j}, t_{k}}
  \end{split}
\end{equation}
with $\mathcal{G}_{s_{0}}=[(s_{0}, s_{1}),...,(s_{0}, s_{K'})]$ denoting the stock
pairs which are among the $K'$-smallest feature RMSPE for stock $s_{0}$.

We repeat the same process for all stocks and we use the same features as in \ref{subsec:temporal_relationship} 
to build another $2 \times K' \times n \times m$ edges.

\bigskip
\noindent
\textbf{Stock Sector}

In finance, each company is classified into a specific sector with Global Industry Classification
Standard\footnote{https://www.msci.com/our-solutions/indexes/gics} (GICS). 
It is shown that the performances of stocks in the same sector 
are often correlated \citep{vardharaj2007sector}. Hence, we simply connect the times of a pair
of stocks if they belong to the same sector. This is written as:
\begin{equation}
  \begin{split}
    \forall\ E_{sec}, \forall s_{i} \in E_{sec}, \forall s_{j} \in E_{sec}\ \textrm{and}\ s_{j} \neq s_{i}, \\
    \forall k \in [1, m],\ \textrm{connect} \ node_{s_{i}, t_{k}}\ \textrm{and}\ node_{s_{j}, t_{k}}
  \end{split}
\end{equation}
where $E_{sec}$ is the ensemble of all the stocks in the sector $sec$.

There are four granularities in GICS sector data:
Sector, Industry Group, Industry, Sub-Industry.
We can therefore construct four different types of edges with this
GICS sector data. In our experiments, we use the Industry
granularity as it gives a good performance with a reasonable number of edges.
The detail of this choice is discussed in Section \ref{subsubsec:sector_relationship}.

\bigskip
\noindent
\textbf{Supply Chain}

Supply chain describes the supplier-customer relation between companies and it is proved
to be useful in multiple financial tasks such as risk management \citep{yang2020financial}
and performance prediction \citep{chen2021graph}.
We use the supply chain data from Factset\footnote{https://www.factset.com/marketplace/catalog/product/factset-supply-chain-relationships}
to build this graph. We connect two companies if they have a supplier-customer relationship in
the training period. This is described as:

\begin{equation}
  \begin{split}
    \forall (s_{i}, s_{j}) \in \mathcal{G}_{supply}, \forall k \in [1, m]\\
    \textrm{connect} \ node_{s_{i}, t_{k}}\ \textrm{and}\ node_{s_{j}, t_{k}}
  \end{split}
\end{equation}
where $\mathcal{G}_{supply}$ is the ensemble of all the supplier-customer relations among the stocks.

\bigskip
\noindent
We show a detailed statistics of each type of relationship we built in Table \ref{tab:edge_stats}.
In addition to using these relations separately, we can also join these relations by simply
putting all the edges together in the same graph. We will show that combining the edges
can help improve the result in Section \ref{sec:experiment_results}.

\begin{table}[htbp]
  \small
  \centering
    \begin{tabular}{ccc}
    \toprule
    Type  & Relation & \textbf{\# edges} \\
    \midrule
    Temporal & Feature Corr & 8.36M \\
    \midrule
    \multirow{3}[2]{*}{Cross-sectional} & Feature Corr & 8.21M \\
          & Sector & 47.86M \\
          & Supply Chain & 22.22M \\
    \midrule
    \multicolumn{2}{c}{Total} & 76.33M \\
    \bottomrule
    \end{tabular}%
    \caption{The number of edges in each relationship we built. 
    The total number denotes the number of all the edges combined. It is
    not exactly the sum of all individual edge counts since there are duplicated edges.
    These numbers are based on the
    nodes in the training set for $\Delta T = 600$.}
    \label{tab:edge_stats}%
\end{table}%

\subsection{Baseline Models}

In order to prove the effectiveness of our model structure,
we also include the performance of some other widely used models as our benchmarks.

\begin{itemize}
  \item {\tt Na\"ive Guess}: We simply use the realized volatility between $t-\Delta T'$ and $t$ to predict the target. 
  It is written as $\widehat{RV_{s, t, \Delta T}} = RV_{s, t-\Delta T', \Delta T'}$.
  \item {\tt HAR-RV}:  Heterogeneous AutoRegressive model of Realized Volatility. A simple but effective realized volatility prediction model proposed by \citet{corsi2009simple}.
  \item {\tt LightGBM}: A gradient boosting decision tree model introduced by \citet{ke2017lightgbm}. It is proven to be highly effective on tabular data.
  \item {\tt MLP}: Multi-Layer Perception network \citep{rumelhart1985learning}. We build a fully connected neural network with three layers, which have 128, 64, 32
  hidden units respectively.
  \item {\tt TabNet}: A neural network proposed by \citet{arik2020tabnet} which specializes in dealing with tabular data.
  \item {\tt Vanilla GTN-VF}: Our Graph Transformer Network for Volatility Forecasting trained without any relationship information.
\end{itemize}

In addition, we show the model performance with different relations individually to demonstrate how different types of relational data
help improve the result compared with Vanilla GTN-VF and other benchmarks.

\subsection{Experiment Results}
\label{sec:experiment_results}

\begin{table*}[h]
  \centering
    \begin{tabular}{ccccccc}
    \toprule
          & \multicolumn{2}{c}{$\Delta T=600$} & \multicolumn{2}{c}{$\Delta T=1200$} & \multicolumn{2}{c}{$\Delta T=1800$} \\
    \midrule
    Model & val   & test  & val   & test  & val   & test \\
    \midrule
    Na\"ive Guess & 0.2911 & 0.2834 & 0.2650 & 0.2628 & 0.2296 & 0.2364 \\
    HAR-RV & 0.2684 & 0.2612 & 0.2149 & 0.2061 & 0.1968 & 0.1939 \\
    LightGBM & 0.2583 & 0.2492 & 0.2414 & 0.2035 & 0.2349 & 0.1963 \\
    MLP   & 0.2431 & 0.2514 & 0.2200 & 0.2308 & 0.2270 & 0.1999 \\
    TabNet & 0.2517 & 0.2478 & 0.2212 & 0.1996 & 0.2204 & 0.2019 \\
    \midrule
    Vanilla GTN-VF & 0.2457 & 0.2498 & 0.2229 & 0.2251 & 0.2092 & 0.2160 \\
    GTN-VF Cross FC & 0.2414 & 0.2382 & 0.2162 & 0.2196 & 0.2066 & 0.2046 \\
    GTN-VF Temp FC & 0.2326 & 0.2358 & 0.1974 & 0.1921 & 0.1896 & 0.1853 \\
    GTN-VF Cross Sector & 0.2406 & 0.2422 & 0.2067 & 0.2248 & 0.2071 & 0.2091 \\
    GTN-VF Cross Supply Chain & 0.2430 & 0.2411 & 0.2105 & 0.2244 & 0.2057 & 0.2000 \\
    GTN-VF Cross FC + Temp FC & 0.2326 & 0.2306 & 0.1936 & 0.1917 & 0.1848 & 0.1802 \\
    GTN-VF & \textbf{0.2314} & \textbf{0.2287} & \textbf{0.1916} & \textbf{0.1892} & \textbf{0.1809} & \textbf{0.1798} \\
    \bottomrule
    \end{tabular}%
    \caption{RMSPE values of all the models on both validation set and test set. In addition to
    the baseline models, we also include the individual performance of the GTN-VF with four relations (Table \ref{tab:edge_stats}), i.e.,
    Cross-sectional Feature Correlation (GTN-VF Cross FC), Temporal Feature Correlation (GTN-VF Temp FC),
    Cross-sectional Sector relationship (GTN-VF Sector) and Cross-sectional Supply Chain relationship (GTN-VF Cross Supply Chain).
    The full GTN-VF includes all four types of relations.
    }
  \label{tab:results}%
\end{table*}%

In our short-term realized volatility forecasting task, 
we use a 3-layer ($L=3$) Graph Transformer Network with 8 heads ($C=8$).
All three layers have 128 channels ($d_{1}=d_{2}=d_{3}=128$).
We embed our categorical features into a 32-dimension vector ($k_{cat}=32$).

The detailed results of our experiments are shown in Table \ref{tab:results}.
We can see that our full GTN-VF model with all four types of relational information 
outperforms all baseline models and each type of relational information individually
on all prediction horizons. The improvement is significant.
In average, we gain 6\% in RMSPE compared with Na\"ive Guess and 2\% compared with
the best baseline model TabNet on test set. It proves that our GTN model structure and
relation building methods are effective.

In terms of individual relationship, all relations are useful since they all show improvement
compared with the vanilla model.
The temporal feature correlation shows
the most predicting power, contributing 1.2\% RMSPE gain while the Sector relation
only contributes 0.7\% improvement when the window is set to 600 seconds although it has
the largest number of edges. This can be explained by the 'noise' included in this type of information
since we need to connect every two stocks in the same sector although not all of them have significant
connection. However, feature correlation only selects the two most related stocks or times, making it 
more discriminational when building edges. This suggests that if we have the constraint 
on the number of edges in the graph, quality is more important than quantity.
On the other hand, these relations are complementary to each other, adding more relations on top
of existing relations can help improve the prediction if we have enough computing power.

It is also worth noting that when we forecast the realized volatility with a longer forward looking and
backward looking window, the result is better. This is simple because the same volatility jump causes more
volatility changes in shorter forecasting horizon \citep{ma2019harnessing}.

\subsection{Ablation Studies}

\subsubsection{Prediction accuracy and stock liquidity}
\label{subsubsec:accuracy_liquidity}

In general, it is more difficult to have a good prediction on less liquid stocks
because there are fewer market participants for them. One sudden change in quote or trade
can cause significant volatility jump, which is difficult to foresee. We analyze the result
to understand where the improvement comes from.

\begin{figure}[]
  \includegraphics[width=\columnwidth]{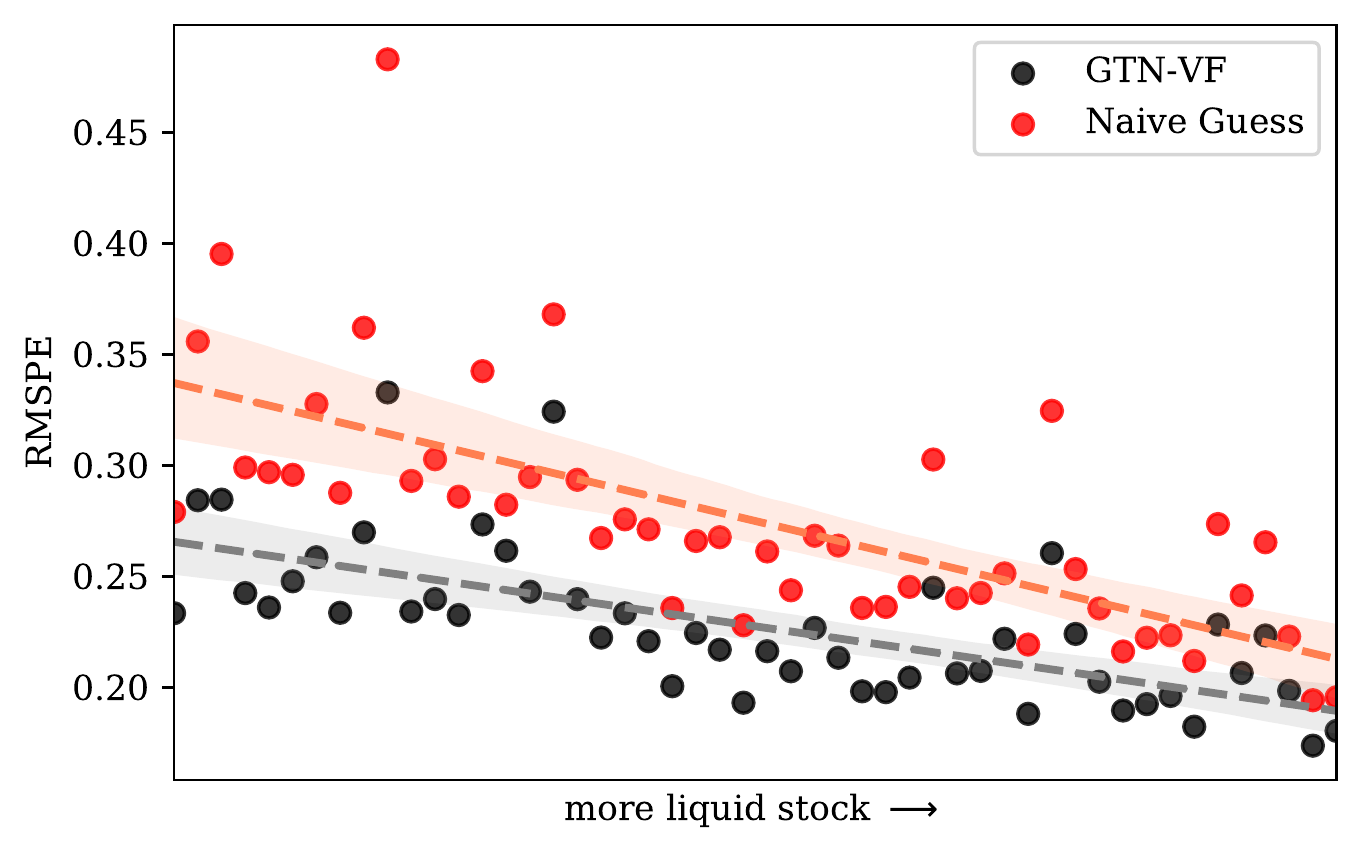}
  \caption{The relationship between stock liquidity and prediction RMSPE.
  The dots denote the RMSPE for the buckets and the dashed lines are the trend lines
  calculated with linear regression.
  This figure is based on test result and $\Delta T=600$.}
  \label{fig:liquidity}
\end{figure}

We first split the stocks into 50 buckets according to their average daily turnover,
which represents the liquidity of a stock. We calculate the RMSPE in each bucket for
both Na\"ive Guess and GTN-VF. The result is illustrated in Figure \ref{fig:liquidity}.

First, we notice that more liquid stocks have smaller RMSPE for both models, which is intuitive.
The prediction for the most liquid stocks is 5\% better than the least liquid stocks in terms of RMSPE, which is 
a large margin in realized volatility forecasting.
We can also see that our graph based model has more improvement on the less liquid stocks 
(around 8\% for more liquid stocks and 2\% for less liquid stocks), although
it is more effective than Na\"ive Guess on all scenarios.

\subsubsection{Prediction accuracy and node connection}

\begin{figure}[h]
  \includegraphics[width=\columnwidth]{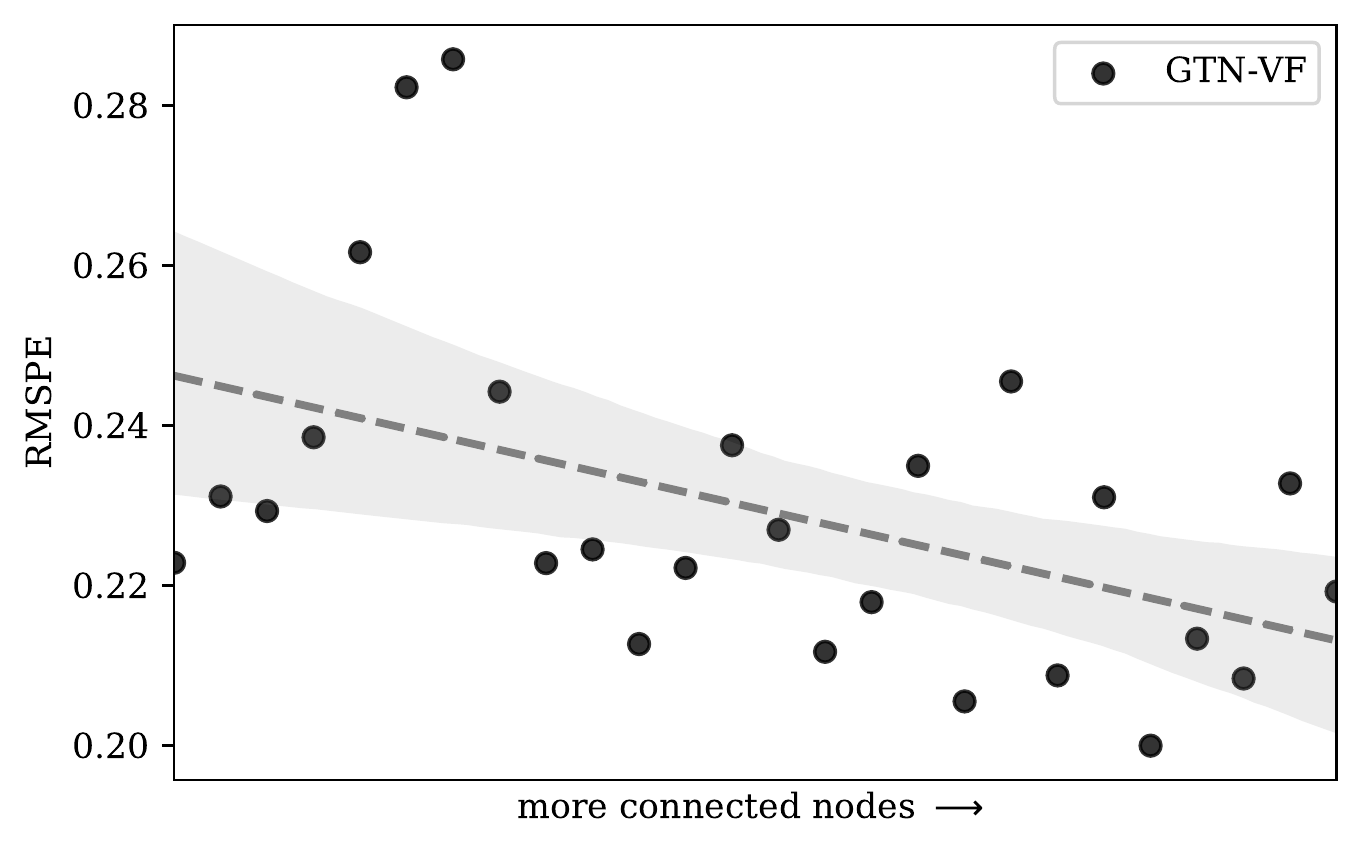}
  \caption{The relationship between node connection and RMSPE.
  This figure is based on test result and $\Delta T=600$.}
  \label{fig:node_connection}
\end{figure}

We also investigate how our model performs on different nodes.
We use the same approach in Section \ref{subsubsec:accuracy_liquidity} by splitting
nodes into buckets according to the number of edges connected to each node.
We can see from Figure \ref{fig:node_connection} that more connected nodes
usually have better RMSPE result, with a 2\% difference between the most connected
and the least connected. This can be explained by the fact that mode connected nodes
make decision based on more information from their neighbor, while the nodes with fewer
or no connections can only rely on the information from themselves. This phenomenon 
proves again the effectiveness of our graph based method.

\subsubsection{Sector Relationship}
\label{subsubsec:sector_relationship}

As introduced in Section \ref{subsubsec:cross_sectional_relationship}, there are four granularities
in the GICS sector data. We run different experiments to evaluate the performance of each
type of sector relationship. The result is shown in Table \ref{tab:sector_relationship}.

\begin{table}[htbp]
  \centering
  \begin{tabular}{ccc}
    \toprule
    Granularity & \# edges & test RMSPE \\
    \midrule
    Sector & 178.7M & N.A. \\
    IndustryGroup & 83.39M & 0.2578 \\
    Industry & 47.86M & 0.2422 \\
    SubIndustry & 19.88M & 0.2441 \\
    \bottomrule
  \end{tabular}%
  \caption{Test RMSPE with different granularities of GICS sector.
  The result for Sector is not available as the number of edges is too
  big to fit in the memory.}
  \label{tab:sector_relationship}%
\end{table}%

We observe that Industry shows the best performance with a modest number of edges.
If we add more edges, such as IndustryGroup, the RMSPE decreases since the relations
among stocks are less meaningful. For the Sector granularity, we are not even able to
obtain a result since the number of edges exceeds the memory limit. Hence, in our final model
combining multiple sources of relations, we choose Industry as our sector relationship.

\section{Conclusion}

We forecast the short-term realized volatility in a multivariate approach.
We design a graph based neural network: Graph Transformer Network for Volatility Forecasting
which incorporates both features from LOB data and relationship among stocks from different sources.
Through extensive experiments on around 500 stocks, we prove the that our method outperforms
other baseline models, both univariate and multivariate. In addition, the model structure allows
to combine an unlimited number of relations, the study of the effectiveness of other relational data
is open for future researches.

\section*{Acknowledgments}

The authors gratefully acknowledge the financial support of the Chaire 
\textit{Machine Learning \& Systematic Methods in Finance}.



\bibliography{paper}
\bibliographystyle{paper}

\newpage

\appendix
\section{List of node features}
\label{sec:list_node_features}

We introduce the features we use in our experiments.

In each bucket, we have $\Delta T'$ lines as training data. We first calculate an indicator
for each line, we then aggregate these indicators in the same bucket with an aggregation function (aggregator).
In such way, we have one value per indicator per aggregator as one feature for the bucket.
The full list of indicators and aggregators are listed in Table \ref{tab:features}.

For some important indicators, we also calculate their progressive features. It means that instead of
applying an aggregator on all the lines, we apply it on the lines between 0 and $\Delta T' / 6$,
$\Delta T' / 3$, $\Delta T' / 2$, $2 \Delta T' / 3$, $ 5 \Delta T' / 6$, $\Delta T'$.
In such way, we have 6 features per indicator per aggregator for an progressive feature.
This is shown in the column Progressive in Table \ref{tab:features}.

We define our aggregation functions as follows. We use $a_{i}$ to denote the $i$-th line in the bucket and $N$ represents the total number of lines.

\begin{itemize}
  \item gini coefficient \citep{gini1921measurement}
  \begin{equation*}
    \frac{\sum_{i}^{N}\sum_{j}^{N}|a_{i} - a_{j}|}{2N^{2}\overline{a}}
  \end{equation*}
  \item percentage difference
  \begin{equation*}
    \frac{\sum_{i}^{N} \mathbf{1}_{a_{i} \neq a_{i-1}}}{N}
  \end{equation*}
  \item realized volatility (Equation \ref{eq:realized_volatility})
  \begin{equation*}
    \sqrt{\frac{1}{N} \sum_{i}^{N} a_{i}^{2}}
  \end{equation*}
  \item percentage greater than mean
  \begin{equation*}
    \frac{\sum_{i}^{N} \mathbf{1}_{a_{i} > \overline{a}}}{N}
  \end{equation*}
  \item percentage greater than zero
  \begin{equation*}
    \frac{\sum_{i}^{N} \mathbf{1}_{a_{i} > 0}}{N}
  \end{equation*}
  \item median deviation
  \begin{equation*}
    median(|a_{i} - \overline{a}|)
  \end{equation*}
  \item energy
  \begin{equation*}
    \frac{1}{N} \sum_{i}^{N} a_{i}^{2}
  \end{equation*}
  \item InterQuartile Range (IQR)
  \begin{equation*}
    Q_{75}(a) - Q_{25}(a)
  \end{equation*}
  where $Q_{i}(a)$ denotes the $i$-th percentile value for the series $a$.
\end{itemize}

\begin{table*}[t!]
  \centering
  \begin{tabular}{cccccc}
    \toprule
    Type  & Notation & Description & Aggregators & Progressive & Count \\
    \midrule
    \multirow{14}[2]{*}{Quote} & \multirow{2}[1]{*}{$\frac{P_{b}^{1} V_{a}^{1} + P_{a}^{1} V_{b}^{1}}{V_{a}^{1} + V_{b}^{1}}$} & \multirow{2}[1]{*}{WAP} & mean, std, gini & \multirow{2}[1]{*}{N} & \multirow{2}[1]{*}{5} \\
    &       &       &  mean of first 100, mean of last 100 &       &  \\
    & $P_{a}^{1}$ & Ask Price & \% difference & N     & 1 \\
    & $P_{b}^{1}$ & Bid Price & \% difference & N     & 1 \\
    & $\frac{P_{a}^{1} - P_{b}^{1}}{P_{a}^{1} + P_{b}^{1}}$ & Price Relative Spread & mean, std, gini & N     & 3 \\
    & $WAP - P_{b}^{1}$ & WAP bid difference & mean, std, gini & N     & 3 \\
    & $log(\frac{WAP_{i}}{WAP_{i-1}})$ & Return & realized volatility & Y     & 6 \\
    & $log(\frac{WAP_{i}}{WAP_{i-1}})^{2}$ & Squared Return & std, gini & N     & 2 \\
    & $\frac{V_{a}^{1} - V_{b}^{1}}{V_{a}^{1} + V_{b}^{1}}$ & Size Relative Spread & mean, std, gini & N     & 3 \\
    & $V_{a}^{1}$ & Ask Size & \% difference & N     & 1 \\
    & $V_{b}^{1}$ & Bid Size & \% difference & N     & 1 \\
    & $V_{a}^{1} / \overline{V_{a}^{1}}$ & Normalized Ask Size & mean, std, gini & N     & 3 \\
    & $V_{a}^{1} + V_{b}^{1}$ & Total Size & sum, max & N     & 2 \\
    & $|V_{a}^{1} - V_{b}^{1}|$ & Size Imbalance & sum, max & N     & 2 \\
    \midrule
    \multirow{10}[2]{*}{Trade} & \multirow{2}[1]{*}{$P_{t}$} & \multirow{2}[1]{*}{Price} & \% greater than mean, \% less than mean & \multirow{2}[1]{*}{N} & \multirow{2}[1]{*}{5} \\
    &       &       & median diviation, energy, IQR &       &  \\
    & \multirow{2}[0]{*}{$log(\frac{P_{t, i}}{P_{t, i-1}})$} & \multirow{2}[0]{*}{Return} & realized volatility & Y     & 6 \\
    &       &       & \% greater than 0, \% less than 0 & N     & 2 \\
    & $log(\frac{P_{t, i}}{P_{t, i-1}})^{2}$ & Squared Return & std, gini & N & 2 \\
    & \multirow{2}[0]{*}{$V_{t}$} & \multirow{2}[0]{*}{Size} & sum   & Y     & 6 \\
    &       &       & max, median diviation, energy, IQR & N     & 4 \\
    & $t$ & Seconds & count & Y     & 6 \\
    & \multirow{2}[0]{*}{$N_{t}$} & \multirow{2}[0]{*}{Order Count} & sum   & Y     & 6 \\
    &       &       & max   & N     & 1 \\
    & $P_{t} \times V_{t}$ & Amount & sum, max & N     & 2 \\
          \midrule
          \multicolumn{2}{c}{Total} &       &       &       & 73 \\
          \bottomrule
    \end{tabular}%
  \caption{The list of features built from LOB data.}
  \label{tab:features}%
\end{table*}%

\end{document}